\documentclass[preprint,
superscriptaddress,
nofootinbib,
 amsmath,amssymb,
 aps,
 prd,
10.5pt]{revtex4-2}

\usepackage{physics}
\usepackage{booktabs}
\usepackage{graphicx}
\usepackage{dcolumn}
\usepackage{bm}
\usepackage{siunitx}
\usepackage{hyperref}
\usepackage{cleveref}
\usepackage{newtxtext}
\usepackage{newtxmath}

\begin{document}


\title{New constraints on axion with gamma-ray observations of the Crab Nebula}

%
%

\author{Kazunori Kohri}
\affiliation{Division of Science, National Astronomical Observatory of Japan (NAOJ), 2-21-1 Osawa, Mitaka, Tokyo 181-8588, Japan}
\affiliation{School of Physical Sciences, Graduate University for Advanced Studies (SOKENDAI), 2-21-1 Osawa, Mitaka, Tokyo 181-8588, Japan}
\affiliation{Theory Center, IPNS, High Energy Accelerator Research Organization (KEK), 1-1 Oho, Tsukuba, Ibaraki 305-0801, Japan}
\affiliation{Kavli IPMU (WPI), UTIAS, The University of Tokyo, Kashiwa, Chiba 277-8583, Japan}
\affiliation{School of High Energy Accelerator Science, Graduate University for Advanced Studies (SOKENDAI), 1-1 Oho, Tsukuba, Ibaraki 305-0801, Japan}
\affiliation{Department of Astronomy, The University of Tokyo, Bunkyo-ku, Hongo, Tokyo 113-0033, Japan}

\author{Haruki Takahashi}
\affiliation{Theory Center, IPNS, High Energy Accelerator Research Organization (KEK), 1-1 Oho, Tsukuba, Ibaraki 305-0801, Japan}
\affiliation{School of High Energy Accelerator Science, Graduate University for Advanced Studies (SOKENDAI), 1-1 Oho, Tsukuba, Ibaraki 305-0801, Japan}

\date{\today}

%
%

\begin{abstract}
In this paper, we derive the upper bounds on the coupling of axion-like particles (ALPs) with photon as a function of the mass by considering axion-photon conversion in the Crab Nebula.  Previous studies have not considered the influence of the magnetic field within the Crab Nebula. The magnetic field plays a crucial role through the Synchrotron Self-Compton (SSC) process, in which high-energy electrons produce synchrotron radiation that is subsequently up-scattered by the same electrons via inverse Compton scattering to generate gamma rays. Therefore, neglecting the magnetic field in modeling leads to theoretical inconsistencies.  In this work, we investigate the significance of the magnetic field effect and demonstrate that even differences in magnetic field modeling can substantially alter the conversion probability. We thus, for the first time, point out that proper consideration of the magnetic field is essential in ALP searches using gamma rays from the Crab Nebula. The resulting constraints reach up to a coupling of $g_{a\gamma \gamma} \lesssim 1 \times 10^{-11} {\rm GeV}^{-1}$ for ALP masses in the range $10^{-10} {\rm eV} \lesssim m_a \lesssim 10^{-6} {\rm eV}$.
\end{abstract}

\maketitle

\newpage

%
%
%
%

\section{Introduction}
Dark matter is a massive and invisible piece in the Universe. A lot of dark matter candidates have been proposed beyond the Standard Model, such as weakly interacting massive particles (WIMPs), primordial black holes (PBHs), axions, and so on (see e.g.~\cite{Safdi:2022xkm} for a lecture note). Axions are pseudo Nambu-Goldstone bosons appearing after an additional $U(1)_{\mathrm{PQ}}$ symmetry is spontaneously broken. It is a dynamical solution to the strong CP problem in quantum chromodynamics (QCD)~\cite{Peccei:1977hh, Peccei:1977ur} and called QCD axions. QCD axions were soon found to work as dark matter in the Universe~\cite{Weinberg:1977ma, Wilczek:1977pj, Preskill:1982cy, Abbott:1982af, Dine:1982ah}, and models were developed such as KSVZ model~\cite{Kim:1979if, Shifman:1979if} and DFSZ model~\cite{Dine:1981rt, Zhitnitsky:1980tq} (see e.g.~\cite{Marsh:2015xka, Marsh:2023tep, Baryakhtar:2025jwh} for reviews). Axion-like particles (ALPs) are a kind of generalizations of the QCD axions and can be a candidate of dark matters. ALPs mass $m_a$ and its coupling to photons $g_{a\gamma\gamma}$ do not have relations with each other as opposed to the QCD axion whose mass is proportional to the coupling. ALPs appear in models beyond the Standard Model such as grand unified theories, higher-dimensional theories, and string theories~\cite{Svrcek:2006yi, Arvanitaki:2009fg, Arvanitaki:2010sy, Tashiro:2013yea, Foster:2022ajl, Gendler:2023kjt, Alexander:2024nvi, Conlon:2024hgw, Reece:2025thc, Gendler:2024adn, Agrawal:2024ejr} (see e.g.~\cite{Ringwald:2012cu, Chadha-Day:2021szb, Reece:2023czb, Choi:2024ome} for reviews and lectures).

For the axion searches (see e.g.~\cite{Graham:2015ouw, Irastorza:2018dyq} for reviews), photon is an important probe and couples to the axions through the Chern-Simons coupling $- \frac{1}{4} g_{a\gamma\gamma} a F_{\mu\nu} \tilde{F}^{\mu\nu}$ where $a$ is the axion field, $F_{\mu\nu}$ is the strength tensor of electromagnetic fields, and $\tilde{F}^{\mu\nu}$ is its dual. This coupling allows the axions to convert into the photons and vise versa. As proposed by Sikivie~\cite{Sikivie:1983ip}, helioscopes and haloscopes are experiments to look for the axions from the Sun and the axion dark matter in the halo, respectively~\cite{CAST:2007jps, CAST:2008ixs, CAST:2011rjr, CAST:2013bqn, CAST:2024eil, IAXO:2019mpb, IAXO:2020wwp, Hudson:2012ee, Ruz:2024gkl, BOREXINO:2025dbp, ADMX:2019uok, ADMX:2020ote, Adair:2022rtw}. CERN Axion Solar Telescope (CAST)~\cite{CAST:2007jps, CAST:2008ixs, CAST:2011rjr, CAST:2013bqn, CAST:2024eil} is one of the helioscopes to detect the solar axions produced in the core of the Sun through the Primakoff process. The CAST have given upper bounds on the axion-photon coupling $g_{a\gamma\gamma} < 6.5 \times 10^{-11} \, \si{GeV^{-1}}$~\cite{CAST:2024eil}. Axion Dark Matter eXperiment (ADMX)~\cite{ADMX:2019uok, ADMX:2020ote} is a haloscope to look for the dark matter axions in the mass range $2.66$ to $\SI{3.1}{\mu eV}$. Light-shining through a wall (LSW) experiments~\cite{Ehret:2010mh, OSQAR:2015qdv, Isleif:2022ytq} such as Any Light Particle Search (ALPS) have also constrained the axion parameters as well as the helioscopes and the haloscopes.

In addition to these experiments, observations of celestial objects also give us a hint for axions~\cite{Mirizzi:2009nq, Bassan:2010ya, Wouters:2013hua, Payez:2014xsa, Meyer:2016wrm, Giannotti:2017hny, Marsh:2017yvc, Reynolds:2019uqt, Reynes:2021bpe, Calore:2020tjw, Carenza:2020cis, Diamond:2023scc, Noordhuis:2022ljw, Hoof:2022xbe, Todarello:2023ptf, Springmann:2024ret, Xu:2024cof, Cyr:2024sbd, Lella:2022uwi, Lella:2023bfb, Lella:2024dmx, Lella:2024hfk, Manzari:2024jns, Carenza:2019pxu, Pallathadka:2020vwu, Carenza:2023lci, Carenza:2025uib, Oshima:2023csb, Pandey:2024dcd, Sisk-Reynes:2022sqd, Galanti:2022tow, Galanti:2022yxn, Benabou:2024jlj, Fiorillo:2025yzf, Fiorillo:2025gnd} (see~\cite{Caputo:2024oqc} for a recent review). For example, gamma-ray observations are ways to get information of axions~\cite{Mirizzi:2007hr, Hooper:2007bq, DeAngelis:2007dqd, DeAngelis:2007wiw, Hochmuth:2007hk, Simet:2007sa, Sanchez-Conde:2009exi, Bassan:2009gy, Mirizzi:2009aj, Belikov:2010ma, DeAngelis:2011id, Tavecchio:2012um, Horns:2012kw, Meyer:2013pny, HESS:2013udx, Meyer:2014epa, Meyer:2014gta, Dobrynina:2014qba, Meyer:2016wrm, Kartavtsev:2016doq, Meyer:2016xve, Vogel:2017fmc, Kohri:2017ljt, Majumdar:2018sbv, Zhang:2018wpc, Galanti:2018upl, Long:2019nrz, Buehler:2020qsn, Mastrototaro:2022kpt, Dong:2025xgo, Cheng:2020bhr, Davies:2020uxn, Guo:2020kiq, Li:2020pcn, Eckner:2022rwf, Li:2021gxs, Li:2024ivs, Li:2024zst, Gao:2023und, Rojas:2023hcg, Wang:2023okw, Gao:2023dvn, Gao:2024wpn, Pant:2022ibi, Pant:2023lnz, Pant:2023omy, Chen:2024fov, Candon:2025vpv, Pratts:2025llp, Kachelriess:2023fta, Jacobsen:2022swa, Davies:2022wvj, Cecil:2023qxn, Cecil:2023hxq, MAGIC:2024arq, Adams:2025izo, Dekker:2025vcg} (see for recent reviews~\cite{Biteau:2022dtt, Galanti:2022ijh, Calore:2025mjv}) as follows. High energy gamma-rays emitted from a source are absorbed by low-energy background photons, and can be converted to the ALPs and vise versa under magnetic fields in the Universe. The gamma-ray absorption and the ALP-photon conversion can alter the gamma-ray spectra and give constraints on the ALP mass $m_a$ and the ALP-photon coupling $g_{a\gamma\gamma}$. While extra-galactic objects such as blazars and radio galaxies have been used for the ALPs search and given constraints on the ALPs parameters, galactic sources have also been recently used in a similar way (e.g.~\cite{Xia:2018xbt, Liang:2018mqm, Xia:2019yud, Bi:2020ths, Li:2024ivs}), such as the Crab Nebula~\cite{Bi:2020ths, Li:2024ivs}. The Crab Nebula is the brightest pulsar wind nebula in the Milky Way (MW) and powered by relativistic electron-positron wind from the Crab Pulsar at the center. Electromagnetic waves from the Crab Nebula have been observed in broad energy range from radio to gamma-rays~\cite{Macias-Perez:2008uaf, DeLooze:2019upw, Aumont:2009dx, FermiPulsarTimingConsortium:2009kko, Fermi-LAT:2016nkz, HAWC:2019xhp, VERITAS:2013pij, MAGIC:2014izm, Amenomori:2019rjd, LHAASO:2021cbz}. These data sets allowed for developing emission models for the Crab Nebula (e.g.~\cite{Meyer:2010tta, Dirson:2022lnl, HESS:2024fri}).

When we use observations of the gamma rays from some astrophysical sources for the ALPs search, it is important to consider astrophysical models of magnetic fields precisely as far as possible. This is because magnetic fields affect on the ALP-photon conversion and then the constraints with the tgamma-ray observations. In this paper, we investigate how the magnetic fields in the Crab Nebula affect on the ALP-photon conversion. This effect has not been taken into account in the ALPs search with gamma-rays from the Crab Nebula so far. As explained in~\cref{sec:modeling}, we consider a model for the magnetic field discussed in~\cite{HESS:2024fri}. The magnetic field is important not only in the ALP-photon conversion but also in photon emission process in the Crab Nebula, called the synchrotron self-Compton (SSC) process which will also be explained in~\cref{sec:modeling}. We find that this consideration significantly affects on the ALP-photon conversion and then alter the resulting constraint on the ALP parameters $m_a$ and $g_{a\gamma\gamma}$. Finally, we obtain new constraints reaching down to $g_{a\gamma\gamma} \lesssim 1 \times 10^{-11}\,\si{GeV^{-1}}$ at $m_a \simeq 10^{-8}\,\si{eV}$ and exceeding the constraints by CAST in the ALP mass range $10^{-10}\,\si{eV} \lesssim m_a \lesssim 10^{-6}\,\si{eV}$.

This paper is organized as follows. In~\cref{sec:axion-photon}, we first show how we can characterize the effects of the gamma-ray absorption and axion-photon conversion.~\cref{sec:modeling} introduces emission models of electromagnetic waves from the Crab Nebula including the model for magnetic fields around the source, which is important not only for the intrinsic gamma-ray emission but also for the ALP-photon conversions. We next perform statistical analyses based on the developed model and constrain the ALP mass $m_a$ and the ALP-photon coupling $g_{a\gamma\gamma}$, and show that magnetic fields in the Crab Nebula also contribute to the ALP-photon conversions and give new constraints on the ALP parameters.

\newpage

%
%
%
%

\section{Axion-Photon Conversion}
\label{sec:axion-photon}

In this section, we use natural units where $c = \hbar = 1$. The system of ALPs $a$ and photons $A_{\mu}$ is described by the following Lagrangian
\begin{equation}
    \mathcal{L} = - \frac{1}{4} F_{\mu\nu} F^{\mu\nu} + \frac{1}{2} \partial_\mu a \, \partial^\mu a - \frac{1}{2} m_a^2 a^2 - \frac{1}{4} g_{a\gamma} a F_{\mu\nu} \tilde{F}^{\mu\nu},
\end{equation}
where $a$ is the ALP field, $m_a$ denotes the ALP mass and $g_{a\gamma}$ is the ALP-photon coupling. $F_{\mu\nu}$ and $\tilde{F}_{\mu\nu}$ are the electromagnetic field strength tensor and its dual, respectively. ALPs and photons are mixing with each other through the  ALP-photon coupling, which can be written as $g_{a\gamma} a \bm{E} \cdot \bm{B}$ in terms of the electric field $\bm{E}$ and magnetic field $\bm{B}$. ALP-photon beams with a monochromatic energy $E$ are propagating along $x_3$-direction under magnetic fields.

The propagation can be described by the equation~\cite{Raffelt:1987im}
\begin{equation}
    \left( i \dv{x_3} + E + \mathcal{M} \right) \Psi(x_3) = 0
\label{eq:equation}
\end{equation}
where $\mathcal{M}$ is the mixing matrix and $\Psi(x_3) = (A_1(x_3), A_2(x_3), a(x_3))^T$. Neglecting Faraday rotation effects, the mixing matrix $\mathcal{M}$
\begin{equation}
    \mathcal{M} = \begin{pmatrix}
        \Delta_\perp & 0 & 0 \\
        0 & \Delta_\parallel & \Delta_{a\gamma} \\
        0 & \Delta_{a\gamma} & \Delta_a
    \end{pmatrix}.
\end{equation}
has several components such as $\Delta_\perp = \Delta_{\mathrm{pl}} + 2 \Delta_{\mathrm{QED}} - i \Gamma / 2$, $\Delta_{\parallel} = \Delta_{\mathrm{pl}} + 7 / 2 \Delta_{\mathrm{QED}} - i \Gamma / 2$, $\Delta_a = - m_a^2 / (2E)$ corresponding to the kinetic term of ALPs, and $\Delta_{a\gamma} = g_{a\gamma} B / 2$ resulting in the ALP-photon conversions. The plasma term $\Delta_{\mathrm{pl}} = - \omega_{\mathrm{pl}}^2 / (2E)$ is arising from the effective photon mass in plasma of astrophysical environments where $\omega_{\mathrm{pl}} = \sqrt{4 \pi n_e e^2 / m_e}$ with the electron density $n_e$, electric charge $e$ and electron mass $m_e$. QED vacuum polarization effects appear in $\Delta_{\mathrm{QED}} = \alpha E / (45 \pi) (B / B_{\mathrm{cr}})^2$ where $\alpha$ is the fine structure constant and $B_{\mathrm{cr}} = m_e^2 / |e| \sim \SI{4.4e13}{G}$. $\Gamma$ characterizes an absorption effect of high-energy gamma-rays as explained below.

High-energy gamma-rays are absorbed through the pair production process where they interact with low-energy background photons and annihilate into electrons and positrons. Cosmic microwave background (CMB) and interstellar radiational fields including star light emission and its dust re-emission~\cite{Moskalenko:2005ng, Vernetto:2016alq, Porter:2018fkx} act as the low-energy background photons. This process occurs when gamma-ray energy is exceeding its threshold value
\begin{equation}
    E_{\mathrm{th}} \sim \frac{4 m_e^2}{\epsilon} \sim \SI{0.5}{TeV} \left( \frac{\epsilon}{\SI{1}{eV}} \right)^{-1}
\end{equation}
with the energy of background photons $\epsilon$. The absorption effect is characterized by
\begin{equation}
    \Gamma = \int d\epsilon \dv{n_{\mathrm{bg}}}{\epsilon} \int_{{-1}}^{{1}} d\cos\theta \frac{1 - \cos \theta}{2} \sigma_{\gamma\gamma}
\end{equation}
where $\mathrm{d}n_{\mathrm{bg}} / \mathrm{d}\epsilon$ is the spectrum of the background photons, $\theta$ is the collision angle between high-energy gamma-rays and background photons. The cross section $\sigma_{\gamma\gamma}$ is written as
\begin{equation}
    \sigma_{\gamma\gamma} = \frac{3}{16} \sigma_T (1 - \beta^2) \left[ 2 \beta (\beta^2 - 2) + (3 - \beta^4) \ln \left( \frac{1 + \beta}{1 - \beta} \right) \right]
\end{equation}
where $\sigma_T$ is the Thomson cross section and $\beta = 1 - 4 m_e^2 / s$ with the Mandelstam variable $s = 2 E \epsilon (1 - \cos \theta)$.

The ALP-photon conversion and gamma-ray absorption affect on the gamma-ray spectrum. We can numerically calculate these effects with a public code \texttt{gammaALPs}~\cite{Meyer:2021pbp} based on an equation of transfer matrices as follows. We first rewrite \cref{eq:equation} as
\begin{equation}
    i \dv{\rho}{x_3} = [\rho, \mathcal{M}]
\label{eq:von-Neumann}
\end{equation}
in terms of the density matrix $\rho(x_3) = \Psi(x_3) \Psi(x_3)^{\dagger}$. Its solution
\begin{equation}
    \rho(z) = \mathcal{T}(x_3, 0; E) \rho(0) \mathcal{T}^{\dagger}(x_3, 0; E)
\end{equation}
can be obtained with the transfer matrix $\mathcal{T}$ which is the solution of \cref{eq:equation} under initial condition $\mathcal{T}(0, 0; E) = 1$. Transfer matrices can be calculated for each environment corresponding to the Crab Nebula $\mathcal{T}_1$ and the Milky Way (MW) $\mathcal{T}_2$. For the environment in MW, we use Jansson $\&$ Farrar model~\cite{Jansson:2012pc}. Models for the Crab Nebula are taken from~\cite{HESS:2024fri} and explained in \cref{sec:modeling}.

In each environment, we assume that the propagation path can be divided into $N_i \, (i=1, 2)$ domains where astrophysical quantities such as the magnetic field and electron density are constant~\cite{Meyer:2021pbp}. The transfer matrix in total path of ALP-photon beams is given by
\begin{equation}
    \mathcal{T}_{\mathrm{total}} = \prod_{i=1}^{2} \prod_{n=1}^{N_i} \mathcal{T}_i (x_{3,\,[N_i-n+1]}, x_{3,\,[N_i-n]}; E)
\end{equation}
where $x_{3,0}$ is the $x_3$ coordinate closest to the Crab Nebula. We then obtain an expression for the photon survival probability
\begin{equation}
    P_{\gamma\gamma} = \mathrm{Tr} \left( (\rho_{11} + \rho_{22}) \mathcal{T}_{\mathrm{total}} \rho(0) \mathcal{T}_{\mathrm{total}}^{\dagger} \right)
\end{equation}
where $\rho_{11} = \mathrm{diag}(1,0,0), \rho_{22} = \mathrm{diag}(0,1,0)$, and $\rho(0) = \mathrm{diag}(1/2, 1/2, 0)$ for unpolarized gamma-rays.

\section{Modeling of the Crab Nebula}
\label{sec:modeling}

\subsection{Electrons and magnetic fields in the Crab Nebula}
We explain the model for the broadband emission of the Crab Nebula~\cite{HESS:2024fri}. Electrons and positrons are distributed around the Crab Nebula. These spectra are characterized with a power-law by the first-order Fermi acceleration. We refer to both electrons and positrons as electrons below. The accelerated electron distribution in Crab Nebula is radially symmetric and steady in this model, and assumed to have two components: radio electrons and wind electrons. Radio electrons are contributing as synchrotron emission from radio to optical regime. At higher energy, synchrotron emission from wind electrons is dominant. The total electron spectrum $n_e(\gamma, r)$ is characterized as
\begin{equation}
    n_e(\gamma, r) = n_{\mathrm{radio}}(\gamma, r) + n_{\mathrm{wind}}(\gamma, r).
\label{eq:electron}
\end{equation}
with the electron's Lorentz factor $\gamma$ and distance from the center of the Crab Nebula $r$.

For the radio electrons,~\cite{HESS:2024fri} assumed that their spectrum $n_{\mathrm{radio}}(\gamma, r)$ follows a power law from $\gamma_{r, \mathrm{min}}$ and has a super-exponential cutoff around $\gamma_{r, \mathrm{max}}$,
\begin{align}
    n_{\mathrm{radio}}(\gamma, r) &= \frac{n_{r, 0}}{\rho_r^3} \gamma^{- s_r} \exp \left[ - \left( \frac{\gamma}{\gamma_{r, \mathrm{max}}} \right)^{\beta_{\mathrm{min}}} \right] \Theta(\gamma - \gamma_{r, \mathrm{min}}) F_{\mathrm{radio}}(r), \\
    F_{\mathrm{radio}}(r) &= \exp \left( - \frac{r^2}{2 \rho_r^2} \right) \Theta(r - r_s),
\end{align}
where $n_{r, 0}$ is the normalization, $s_r$ is the index, $\beta_{\mathrm{min}}$ is responsible for the super-exponential cutoff, and $\Theta(x)$ is the Heaviside step function. $F_{\mathrm{radio}}(r)$ shows the radial dependence of the electron spectrum and follows a Gaussian function with constant width $\rho_r$. 

Wind electron spectrum $n_{\mathrm{wind}}(\gamma, r)$ is modeled as
\begin{align}
    n_{\mathrm{wind}}(\gamma, r) &= \frac{n_{w, 0}}{\rho_w(\gamma)^3} G(\gamma) \exp \left[ - \left( \frac{\gamma_{w, \mathrm{min}}}{\gamma} \right)^{\beta_{\mathrm{min}}} \right] \exp \left[ - \left( \frac{\gamma}{\gamma_{w, \mathrm{max}}} \right)^{\beta_{\mathrm{max}}} \right] F_{\mathrm{wind}}(\gamma, r), \\
    G(\gamma) &= \left( \frac{\gamma}{\gamma_{w, 1}} \right)^{- s_{w, 1}} \left( \frac{\gamma_{w,1}}{\gamma_{w,2}} \right)^{-s_{w,2}} \left( 1 - \Theta(\gamma - \gamma_{w,1}) \right) \\
    &+ \left( \frac{\gamma}{\gamma_{w,2}} \right)^{-s_{w,2}} \left( \Theta(\gamma - \gamma_{w,1}) - \Theta(\gamma - \gamma_{w,2}) \right) \\
    &+ \left( \frac{\gamma}{\gamma_{w,2}} \right)^{-s_{w,3}} \Theta(\gamma - \gamma_{w,2}), \\
    F_{\mathrm{wind}}(\gamma, r) &= \exp \left( - \frac{r^2}{2 \rho_w(\gamma)^2} \right) \Theta(r - r_s).
\end{align}
The spectrum has super-exponential cutoffs at $\gamma_{w,\mathrm{min}}$ and $\gamma_{w,\mathrm{max}}$. $G(\gamma)$ denotes a double broken power law of the spectrum with the indexes $s_{w,1}, s_{w,2}$ and $s_{w,3}$. The spatial distribution $F_{\mathrm{wind}}(\gamma, r)$ is assumed to follow a Gaussian and also depend on the energy of the wind electrons. Its dependence is given by
\begin{equation}
    \rho_{w}(\gamma) = \rho_{w,0} \left[ \left( \frac{\gamma}{9 \times 10^5} \right)^2 \right]^{-\alpha_w}.
\end{equation}

The radio and wind electrons radiate low-energy photons through synchrotron process under magnetic fields around the Crab Nebula. For a model of magnetic fields near the Crab Nebula, we consider a variable $B$-field model developed in~\cite{Dirson:2022lnl,HESS:2024fri}. In the variable $B$-field model, the magnetic field has a power-law profile with respect to $r$
\begin{equation}
    B(r) = B_0 \left( \frac{r}{r_s} \right)^{-\alpha_b}
\label{eq:magnetic}
\end{equation}
where $\alpha_b\,(\geq 0)$ is the index, $r_s$ is a shock radius, and $B_0$ is the magnetic field strength at $r_{{s}}$.

\subsection{Emission mechanism of the Crab Nebula}

Using the expressions for the distribution of electrons $n_e(\gamma, r)$ [\cref{eq:electron}] and magnetic fields $B(r)$ [\cref{eq:magnetic}], we can write down the spectral volume emissivity $j_{\nu}(\nu, r) \equiv dE / dt\,dV\,d\nu\,d\Omega$. For the synchrotron process, its spectral volume emissivity is expressed as
\begin{equation}
    j_{\nu}^{\mathrm{sync}}(\nu, r) = \frac{1}{4\pi} \frac{\sqrt{3} e^3 B(r)}{m_e c^2} \int_1^{\infty} d\gamma \, n_e(\gamma, r) f_{\mathrm{sync}}\left( \frac{\nu}{\nu_c} \right)
\label{eq:sync emissivity}
\end{equation}
where $c$ is the speed of light and $\nu_c = 3eB\gamma^2 / (4\pi m_e c)$ is the critical frequency. Note that we do not use the natural unit in this \cref{sec:modeling}. The function $f_{\mathrm{sync}}(x)$ is given by
\begin{equation}
    f_{\mathrm{sync}}(x) = \frac{x}{20} \left[ (8 + 3x^2) \left( K_{1/3}(x/2) \right)^2 + x K_{2/3}(x/2) \left( 2 K_{1/3}(x/2) - 3x K_{2/3}(x/2) \right) \right]
\end{equation}
where $K_{\xi}$ is the modified Bessel function of kind $\xi$.

Low-energy photons radiated through the synchrotron process can be kicked up by high-energy electrons and become high-energy gamma-rays by inverse-Compton (IC) scattering. This is called a synchrotron self-Compton (SSC) process. The emissivity of the IC scattering
\begin{equation}
    j_{\nu}^{\mathrm{IC}}(\nu, r) = \frac{3\sigma_T h c}{4} \frac{h\nu}{4\pi} \int_1^{\infty} d\gamma \, \frac{n_{\mathrm{el}}(\gamma ,r)}{\gamma^2} \int_0^{\infty} d\epsilon \, f_{\mathrm{IC}}(\nu, \epsilon, \gamma) \frac{n_{\mathrm{seed}}(\epsilon, r)}{\epsilon}
\label{eq:ic emissivity}
\end{equation}
can be calculated with the IC kernel function
\begin{equation}
    f_{\mathrm{IC}}(x) = 2q \ln q + (1 + 2q) (1 - q) + \frac{1}{2} \frac{(\Gamma_{\epsilon} q)^2}{1 + \Gamma_{\epsilon} q} (1 - q)
\end{equation}
where $\sigma_T$ is the Thomson cross section, $h$ is the Planck constant, $\Gamma_{\epsilon} = 4 \epsilon \gamma / (m_e c^2)$ and $q = h \nu / (\Gamma_{\epsilon} (\gamma m_e c^2 - h \nu)$). $n_{\mathrm{seed}}(\epsilon, r)$ denotes seed photon's density where $\epsilon$ is the energy of seed photons.

While synchrotron photons are important as seed photons of the SSC process, cosmic microwave background (CMB) and dust emissions are also components of seed photons. CMB photons act as not only one of seed photons but also prevent especially PeV gamma rays from propagating for long distance in the Universe~\cite{Kohri:2012tq} as explained in \cref{sec:axion-photon}, and its density is
\begin{equation}
    n_{\mathrm{seed}}^{\mathrm{CMB}} = \frac{4\pi}{h c} \frac{B_{\nu}(T_{\mathrm{CMB}})}{h \nu}.
\end{equation}
where $B_{\nu}(T)$ is the intensity of the black body at temperature $T$. Dust emission was introduced to explain the observational data in the radio and optical ranges (e.g.~\cite{DeLooze:2019upw}).~\cite{HESS:2024fri} used a model where the dust emission is originated from a mixture of two populations consisting of amorphous carbon dust grains with different temperatures $T_i \,(i=1,2)$ and their masses $M_i \,(i=1,2)$. The dust populations are within a shell whose inner and outer radii are $r_{\mathrm{in}}$ and $r_{\mathrm{out}}$, respectively. The emissivity of the dust is given by
\begin{equation}
    j_{\nu}^{\mathrm{dust}}(\nu, r) = \frac{3 \kappa [\Theta(r - r_{\mathrm{in}}) - \Theta(r - r_{\mathrm{out}})]}{4\pi (r_{\mathrm{out}}^3 - r_{\mathrm{in}}^3)} \sum_{i=1,2} M_i B_{\nu}(T_i)
\end{equation}
where $\kappa$ is the absorption coefficient dependent on the wavelength $\lambda$
\begin{equation}
    \kappa = 2.15 \times 10^{-4} \, \si{cm^2} \si{g^{-1}} \left( \frac{\lambda}{\mu \si{m}} \right)^{-1.3}.
\end{equation}

We can calculate the density of synchrotron and dust emissions as seed photons of SSC process
\begin{equation}
    n_{\mathrm{seed}}^j(\epsilon, r) = \frac{r_0}{2hc} \frac{4\pi}{\epsilon} \int_{r_s/r_0}^1 dy \, \frac{y}{x} \ln \frac{x + y}{|x + y|} j_{\nu}^t(\nu, r_0 y), \hspace{10pt} (j=\mathrm{sync, dust}).
\end{equation}
where $r_0 = \SI{2.0}{pc}$ is the radius of the Crab Nebula. The total density of seed photons is expressed by
\begin{equation}
    n_{\mathrm{seed}}(\epsilon, r) = \sum_k n_{\mathrm{seed}}^k(\epsilon, r), \hspace{10pt} (k = \mathrm{sync, dust, CMB})
\end{equation}
and used to calculate the IC emissivity in \cref{eq:ic emissivity}. The total luminosity $L_{\nu}$ can be calculated by integrating the emissivity in the SSC process over volume $dV$ and solid angle $d\Omega$ as
\begin{equation}
    L_{\nu} = \oint d\Omega \int dV j_{\nu}(\nu, r) = (4 \pi)^2 \int_{r_s}^{r_0} dr \, r^2 \left[ j_{\nu}^{\mathrm{sync}}(\nu, r) + j_{\nu}^{\mathrm{IC}}(\nu, r) \right],
\end{equation}
using \cref{eq:sync emissivity} and \cref{eq:ic emissivity}.

\subsection{Axion-photon conversion in the Crab Nebula}

\begin{table}[t]
    \centering
    {\setlength{\tabcolsep}{14pt} 
    \begin{tabular}{l|c}
        \toprule
        Parameter & variable $B$-field model  \\
        \midrule
        $B_{{0}}$ [\si{\mu G}] & 256.4 \\
        $\alpha_b$ & 0.4691 \\
        $r_s$ [\si{pc}] & 0.13 \\
    \bottomrule
    \end{tabular}
    }
    \caption{Parameters of magnetic fields in the variable $B$-field model taken from~\cite{HESS:2024fri}.}
    \label{tab:B parameters}
\end{table}

This model~\cite{HESS:2024fri} enables us to numerically calculate the flux of high-energy gamma-rays. We can obtain parameters for magnetic fields based on this model shown in~\cref{tab:B parameters}. In the model of the variable magnetic field, we adopt  the best-fit value $\alpha_b = 0.4691$ to fit the full multi-wavelength spectrum without the ALPs, which gives the conservative upper bound on the model parameters of the ALPs. However, the numerical calculations are consuming a lot of time and computer resources to fit the broad emission from radio to gamma-rays. Since we would like to see how ALPs affect the spectrum of high-energy gamma-rays only, we  focus on the gamma-ray spectrum in GeV $\sim$ PeV energy range. A similar approach was taken in the case of ALP search using gamma-rays from a blazar~\cite{Gao:2024wpn}. Fermi-LAT has observed the gamma-rays consisted of synchrotron and IC emissions around GeV scale~\cite{Buehler:2011aa, Arakawa:2020yfc}, and LHAASO reported TeV and PeV gamma-rays in IC part~\cite{LHAASO:2021cbz}. It is known that the gamma-ray spectrum in the energy scale can be well described by the following function~\cite{Arakawa:2020yfc, LHAASO:2021cbz}
\begin{align}
    \Phi_{\mathrm{int}} &= N^0_{\mathrm{sync}} \left( \frac{E}{\SI{100}{MeV}} \right)^{-\Gamma_{\mathrm{sync}}} + N^0_{\mathrm{IC}} \left( \frac{E}{\SI{1}{GeV}} \right)^{- \alpha - \beta \ln (E / \SI{1}{GeV})}
\label{eq:intrinsic spectrum}
\end{align}
where $E = h\nu$ is the energy of gamma-rays{, and \cref{eq:intrinsic spectrum} will be used in the following analysis}. A power-law spectrum with the index $\Gamma_{\mathrm{sync}}$ was used for the synchrotron part~\cite{Arakawa:2020yfc} while IC emission was explained by using a log-parabola spectrum whose shape is characterized by $\alpha$ and $\beta$~\cite{Arakawa:2020yfc, LHAASO:2021cbz}. $N^0_{\mathrm{sync}}$ and $N^0_{\mathrm{IC}}$ are the normalization constants for synchrotron and IC emissions, respectively. {Here, $\Gamma_{\mathrm{sync}}, \alpha, \beta, N^0_{\mathrm{sync}}$ and $N^0_{\mathrm{IC}}$ are free parameters to be fitted. If we denote the spectrum of the primary electron number-density as $dn_e/dE
\propto E^{-p_e}$, the spectrum of the synchrotron emission is represented by $E^2 dN/dE
\propto E^{-(p_e-3)/2}$. The spectrum of the standard IC is also the same:
$E^2 dN/dE \propto E^{-(p_e-3)/2}$ at low energies.  However, in the
case of the SSC, there also exists a higher energy range where the spectrum
becomes flatter, e.g., $E^2 dN/dE \propto
E^{-(p_e-2)/2}$~\cite{Nakar:2009er,Kohri:2012tq}.  With the parameters
used in this paper therefore, we have $\Gamma_{\rm
sync} \sim \alpha \sim (p_e - 3)/2$ in most of the energy regions. On the other hand,  there 
surely exists an energy region where $\alpha \sim (p_e-2)/2$ at a higher energy. This 
serves as evidence for the SSC. In fact, this feature of the spectrum for the SSC has been
confirmed in the Crab Nebula (e.g., see Ref.~\cite{Kohri:2012tq}) and can be used to constrain the parameters of the ALPs}

As explained in \cref{sec:axion-photon}, the intrinsic spectrum of gamma-rays $\Phi_{\mathrm{int}}$ emitted from the source undergoes the ALP-photon conversion and gamma-ray absorption during its propagation to Earth. We can numerically calculate the model flux $\Phi_{\mathrm{model}}$ as
\begin{equation}
    \Phi_{\mathrm{model}} = P_{\gamma\gamma} \, \Phi_{\mathrm{int}}.
\end{equation}
The photon survival probability $P_{\gamma\gamma}$ determines how the gamma-ray spectrum can be distorted by the effects of the ALP-photon conversion and gamma-ray absorption. Note that we took into account the effects of magnetic fields of the Crab Nebula on ALP-photon conversions. Since we have considered the model for magnetic fields in the Crab Nebula, this can alter constraints on the ALP mass $m_a$ and coupling to photons $g_{a\gamma\gamma}$ as discussed in \cref{sec:analysis}.

\section{Analysis and Results}
\label{sec:analysis}

We minimize chi-squared function
\begin{equation}
    \chi^2 = \sum_i \frac{(\Phi^i_{\mathrm{model}} - \Phi^i_{\mathrm{data}})^2}{\sigma_i^2}
\end{equation}
to get the best-fit spectrum where $\Phi^i_{\mathrm{model}}$, $\Phi^i_{\mathrm{data}}$ and $\sigma_i$ are the predicted flux, observed one and its error bar, respectively. Each $i$ corresponds to a data point from experiments. The test statistic (TS) are defined as $\mathrm{TS} = \chi^2_{\mathrm{w/\,ALP}} - \chi^2_{\mathrm{w/o\,ALP}}$. We cannot apply the Wilks' theorem~\cite{Wilks:1938dza} because of non-linear effects on the gamma-ray spectrum in the ALP-photon system~\cite{Fermi-LAT:2016nkz}, and should get a TS distribution with Monte Carlo (MC) simulations as performed in~\cite{Li:2020pcn, Li:2021gxs, Li:2024zst, Gao:2023dvn}. 500 mock data sets of the photon flux are randomly produced with Gaussian sampling. Each data point has the mean value and error bar which are set to be the predicted flux based on the best-fit spectrum and the experimental error, respectively. Optimizing the model flux to these mock data sets reproduces the TS distribution under null hypothesis. This distribution is considered to be the approximation of the TS distribution under the ALP hypothesis~\cite{Fermi-LAT:2016nkz} and it turns out to follow a non-central chi-square distribution. Its non-centrality and effective degree of freedom (d.o.f.) are found to be 0.82 and 4.06, respectively. The $95\%$ quantile value is $\Delta \mathrm{TS}_{95\%} = 11.41$. We can exclude the parameter region where the best-fit chi-square under the ALP hypothesis $\hat{\chi}^2_{\mathrm{ALP}}$ exceeds $\hat{\chi}^2_{\mathrm{min}} + \Delta \mathrm{TS}_{95\%}$ at $95\%$ confidence level (C.L.). Here, $\hat{\chi}^2_{\mathrm{min}}$ is the global best-fit chi-square for the observational data in the scanned parameter space spanning $m_a \in [10^{-11}, 10^{-5}]\,\si{eV}$, $g_{a\gamma} \in [10^{-12},\,5 \times 10^{-10}]\,\si{GeV^{-1}}$.

\begin{figure}[p]
    \centering
    \includegraphics[width=0.95\linewidth]{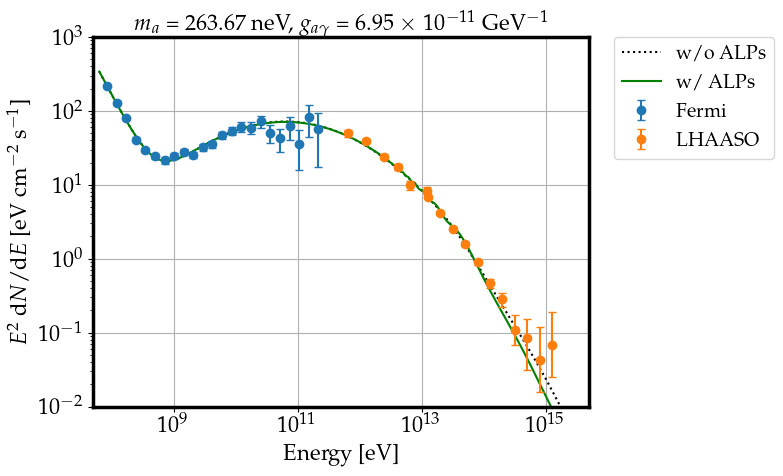}

    \includegraphics[width=0.95\linewidth]{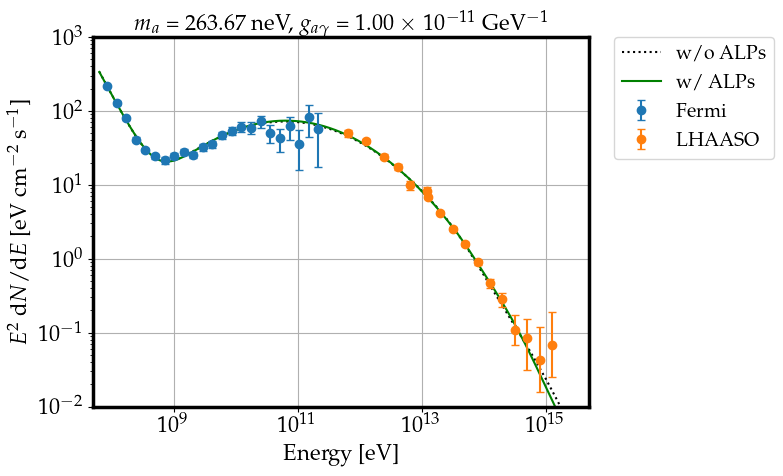}
    \caption{Spectral energy distribution without and with ALPs in some parameter sets. While the parameter set for the top panel is excluded in \cref{fig:variable-B}, the bottom one is allowed with decreasing the axion-photon coupling $g_{a\gamma\gamma}$.}
    \label{fig:SED 1}
\end{figure}

\begin{figure}[p]
    \centering
    \includegraphics[width=0.95\linewidth]{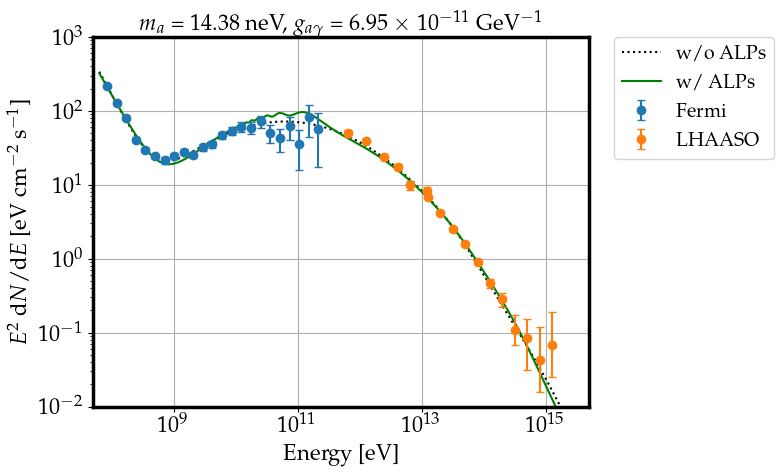}

    \includegraphics[width=0.95\linewidth]{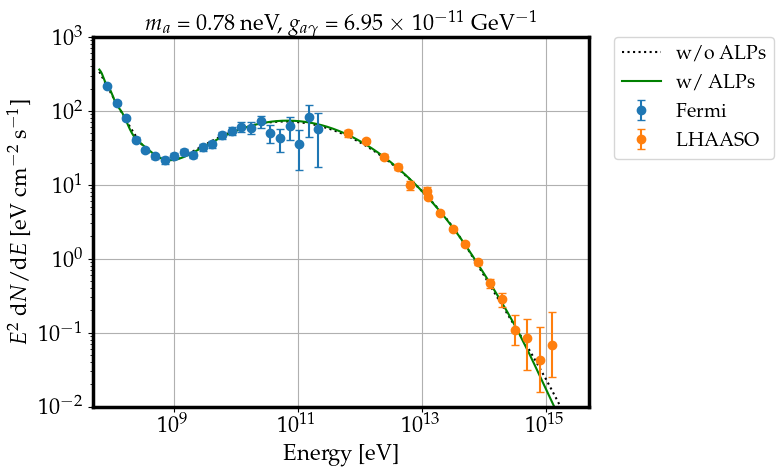}
    \caption{Same as \cref{fig:SED 1} but in different parameter sets. The parameter set corresponding to the top panel is excluded in \cref{fig:variable-B}. With decreasing the axion mass $m_a$, the parameter set for the bottom panel is allowed.}
    \label{fig:SED 2}
\end{figure}

\clearpage

\begin{figure}[p]
    \centering    
    \includegraphics[width=0.95\linewidth]{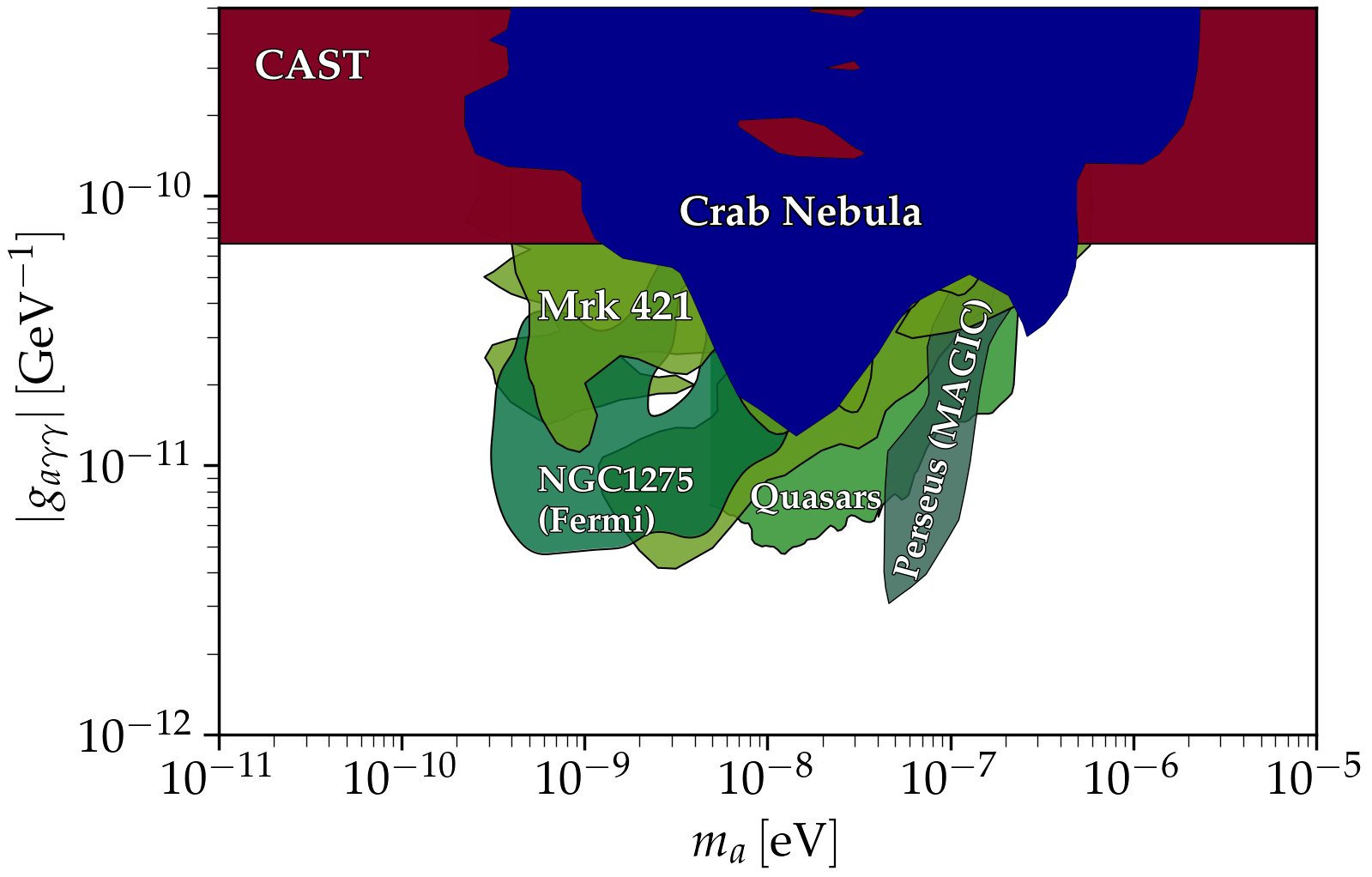}
    \caption{Excluded region at 95$\%$ C.L. is shown as {blue} based on the observations of gamma rays from the Crab Nebula. Red region is excluded by CAST experiments~\cite{CAST:2024eil}. {Green region corresponds to existing constraints (e.g. NGC 1275 in the Perseus galaxy cluster~\cite{Fermi-LAT:2016nkz, MAGIC:2024arq}, Mrk 421~\cite{Li:2020pcn,Li:2021gxs,Li:2024zst,Gao:2024wpn}, quasars~\cite{Davies:2022wvj}, and other constraints in~\cite{AxionLimits} and references therein.).} This figure is produced with \texttt{AxionLimits}~\cite{AxionLimits}.}
    \label{fig:variable-B}
\end{figure}

\clearpage

Following the statistical analysis described above, we use the observational data of the Crab Nebula with Fermi-LAT~\cite{Buehler:2011aa} and LHAASO~\cite{LHAASO:2021cbz} to get constraints on ALP parameters $m_a$ and $g_{a\gamma\gamma}$. In \cref{fig:SED 1,fig:SED 2}, we show these observational datasets and the best-fit spectrum with and without ALPs in some parameter sets, for instance. \cref{fig:variable-B} represents excluded region ($95\%$ C.L.) with high-energy gamma rays from the Crab Nebula as {blue}. CAST has excluded red region~\cite{CAST:2024eil}. As shown in in \cref{fig:variable-B}, the parameter sets for the top panel in \cref{fig:SED 1,fig:SED 2} are excluded while ones for the bottom panel are allowed with decreasing the axion-photon coupling $g_{a\gamma\gamma}$ in \cref{fig:SED 1} or the axion mass $m_a$ in \cref{fig:SED 2}. Taking into account the effects of the magnetic field in the Crab Nebula, the resulting constraint reaches up to $g_{a\gamma\gamma} \lesssim 1 \times 10^{-11} \,\si{GeV^{-1}}$ at $m_a \simeq 10^{-8}\,\si{eV}$, and is exceeding the constraints by CAST in the mass range $10^{-10}\,\si{eV} \lesssim m_a \lesssim 10^{-6} \, \si{eV}$. The shown parameter space which is excluded by the Crab Nebula in this work might be milder than other gamma-ray observations (e.g. NGC 1275 in the Perseus galaxy cluster~\cite{Fermi-LAT:2016nkz, MAGIC:2024arq}, Mrk 421~\cite{Li:2020pcn,Li:2021gxs,Li:2024zst,Gao:2024wpn}, quasars~\cite{Davies:2022wvj}, and other constraints in~\cite{AxionLimits} and references therein.). However, a reliable constraint on the parameters from observations should be the one derived under the most conservative assumptions. One should avoid obtaining overly stringent constraints under intentionally-strong assumptions. Our constraints arise from the fact that the magnetic-field effect appears twice in the SSC process: once in synchrotron radiation and again in the scattering of that radiation via the inverse Compton process. This aspect was overlooked in the previous works. Consequently, even with the most conservative treatment, we obtained unavoidably more stringent constraints. Then, our limit can be regarded to be newly compliment to those other limits.

We can roughly understand the shape of the excluded region as follows. For sizable conversions from ALPs to photons, the gamma-ray energy $E$ and propagation distance $L$ should satisfy
\begin{align}
    E &\gtrsim E^* = \frac{m_a^2}{2g_{a\gamma\gamma}B} \label{eq:condition E} \\
    L &\gtrsim L^* = \frac{2}{g_{a\gamma\gamma}B} \label{eq:condition L}.
\end{align}
$E^*$ and $L^*$ can be written in a suitable unit as~\cite{Kohri:2017ljt}
\begin{align}
    E^* &\sim \frac{\SI{10}{GeV}\,m_{a,\si{neV}}^2}{g_{11} B_{\SI{10}{\mu G}}} \\
    L^* &\sim \frac{\SI{10}{kpc}}{g_{11}B_{\SI{10}{\mu G}}}
\end{align}
with the following notation $m_{a,\si{neV}} \equiv m_a / \si{neV}$, $g_{11} \equiv g_{a\gamma\gamma}/\SI{e-11}{GeV^{-1}}$ and $B_{\SI{10}{\mu G}} \equiv B/\SI{10}{\mu G}$. The region below the green area in \cref{fig:variable-B} is allowed since \cref{eq:condition L} is not satisfied for galactic sources in the MW, corresponding to the upper bounds $g_{a\gamma\gamma} \lesssim 1 \times \SI{e-11}{GeV^{-1}}$. The right region around $m_{a,\si{neV}} \sim 1000$ is not excluded by the condition \cref{eq:condition E} meaning that this condition \cref{eq:condition E} necessitates ultra high-energy gamma rays with energy above PeV scale. We could not conservatively exclude the left region by the following reason. $E^*$ becomes smaller than $\SI{1}{GeV}$ in the parameter space $m_{a,\si{neV}} \lesssim 0.1$ and $1 \lesssim g_{11} \lesssim 50$ and then the ALP-photon conversion is efficient for the energy range observed by Fermi-LAT~\cite{Buehler:2011aa} and LHAASO~\cite{LHAASO:2021cbz}. However, this effect is compensated by the parameters for the intrinsic spectrum and we cannot distinguish the case with and without ALPs. These imply that, roughly speaking, the observed gamma-ray energy correspond to the mass range excluded in this analysis ($0.1 \lesssim m_{a,\si{neV}} \lesssim 1000$).

Here, let us put a comment on statistical analyses. We did not use so-called $\mathrm{CL_s}$ method, which was used in ALP search with gamma-ray observations (e.g.~\cite{Gao:2023dvn, Gao:2024wpn, Li:2024ivs, Adams:2025izo}) as well as high-energy experiments~\cite{Junk:1999kv, Read:2002hq, Lista:2016chp}. It is known that this method allows us to get less optimistic constraints on considering parameters than the traditional frequentist approach when it is difficult to distinguish the null and alternative hypothesis with experiments. However, $\mathrm{CL_s}$ method requires high computational resources because one needs to optimize the model flux with mock data sets for each point in the broad parameter space.~\cite{Gao:2023dvn} compared the $\mathrm{CL_s}$ method with the one used in~\cite{Fermi-LAT:2016nkz, Li:2020pcn, Li:2021gxs, Li:2024zst}, and the authors concluded that the latter is also consistent with the former. This is why we took the latter approach to reduce time for numerical calculations.

\clearpage

\section{Conclusion}
In this paper, we have investigated the importance of magnetic fields in the Crab Nebula for the ALP search with the gamma-rays. The ALP-photon beams experience the gamma-ray absorption and the ALP-photon conversion in their journey from the source. For the models of magnetic fields around the Crab Nebula, we considered the variable $B$-field model~\cite{HESS:2024fri}. It turned out that the magnetic fields in the Crab Nebula affected on how the gamma-ray spectrum is distorted by the ALP-photon conversion (\cref{fig:SED 1,fig:SED 2}). This effect has not been taken into account in ALP search with gamma rays from the Crab Nebula before. By considering the magnetic fields of the Crab Nebula, we obtain new constraints reaching down to $g_{a\gamma\gamma} \lesssim 1 \times 10^{-11}\,\si{GeV^{-1}}$ at $m_a \simeq 10^{-8}\,\si{eV}$ and exceeding the constraints by CAST in the ALP mass range $10^{-10}\,\si{eV} \lesssim m_a \lesssim 10^{-6}\,\si{eV}$. This consideration can be applicable to other galactic sources. The analysis in this paper will be a step forward to more precise constraints on the ALP parameters as briefly mentioned below.

Let us give a comment on the statistical analysis and other model uncertainties. As explained in~\cref{sec:analysis}, we have not used the $\mathrm{CL_s}$ method for saving times for numerical calculations. Although it was shown in~\cite{Gao:2023dvn} that the statistical analysis used in~\cite{Fermi-LAT:2016nkz, Li:2020pcn, Li:2021gxs, Li:2024zst} and this paper gave consistent results compared to the $\mathrm{CL_s}$ method. However, we may have a possibility to obtain more conservative results based on the $\mathrm{CL_s}$ method after increasing computer resources. We have also a comment on model uncertainties of the Crab Nebula. Although the Crab Nebula has been observed for long time and in broad energy range allowing for developing SSC emission models, there still is a little room for adding other contributions to the spectrum of the Crab Nebula. For example,~\cite{Spencer:2025abk} considered a hadronic contribution such as neutral pion decay in addition to the leptonic component in order to fit the gamma-ray spectrum with LHAASO data~\cite{LHAASO:2021cbz} at PeV scale. If we take this possibility into account, we could extend our analysis performed in this paper. These remain as future works.

\clearpage

\begin{acknowledgments}
This work was in part supported by JSPS KAKENHI Grants Nos. JP23KF0289,
JP24K07027 (K.K.) and JP24KJ1153 (H.T.), and MEXT KAKENHI Grants No. JP24H01825 (K.K.).
\end{acknowledgments}

\appendix


\bibliographystyle{apsrev4-2}
\bibliography{alps_crab}

\end{document}